\documentclass{iopart}
\usepackage{epsfig}
\begin{document}

\title[]{Heavy ion collisions and  lattice QCD at finite baryon density}

\author{Krzysztof Redlich\dag\footnote[4]{ redlich@rose.ift.uni.wroc.pl
 }, Frithjof Karsch\ddag\footnote[5]{
 karsch@physik.uni-bielefeld.de
 }  ~and Abdelnasser
  Tawfik\ddag  \footnote[6]{ tawfik@physik.uni-bielefeld.de
 } }

\address{\dag\ Institute of Theoretical Physics University of Wroclaw,
PL-50204 Wroclaw, Poland }

\address{\ddag\ Fakult\"at f\"ur Physik, Universit\"at Bielefeld,
Postfach 100 131, D-33501 Bielefeld, Germany}

\begin{abstract}
 We discuss a relation  between the
 QCD thermodynamics obtained from a statistical analysis of particle production in heavy ion
 collisions at SPS and RHIC energies and recent LGT results at finite chemical
 potential. We
 show that basic thermodynamic properties obtained from the
 phenomenological statistical operator of a hadron
 resonance gas that
 describes particle yields in heavy ion collisions are consistent
 with recent LGT results. We argue that for $T\leq T_c$ the equation of state derived
 from
 Monte--Carlo simulations of two quark--flavor QCD at finite
chemical potential can be well described by a hadron resonance gas
when using the same set of approximations as used in LGT calculations.
We examine the influence of a finite quark mass  on the position
of the deconfinement transition in temperature and chemical potential
plane.
\end{abstract}

\section{Introduction}
The detailed analysis of particle production in heavy ion
collisions has shown that in a broad energy range from SIS through
AGS, SPS up to RHIC particle yields resemble that of chemical
equilibrium population \cite{rev}. At SPS and RHIC the freeze-out
parameters, the  temperature $T_f$ and the energy density
$\epsilon_f$, predicted by the presently used thermal model of
hadron resonance gas, agree well with recent results from the
lattice on critical conditions required for deconfinement
\cite{rev,our}. The above quantitative agreement of freeze-out and
critical parameters suggests that at SPS and RHIC chemical
freeze-out appears in the near vicinity or at the phase boundary
\cite{stach}. If this is indeed the case then the phenomenological
statistical operator of hadron resonance gas $Z_{HG} $ should also
provide a  consistent with LGT  description of QCD thermodynamics
in the confined, hadronic phase \cite{our}. Here we show that the
basic qualitative properties of $Z_{HG} $ resulting from its
dependence on $T$ and baryon chemical potential $\mu_B$ are also
present in recent lattice results of two flavor QCD at finite
chemical potential.
In  addition, when imposing the fixed energy density condition for
deconfinement the hadron resonance gas partition function $Z_{HG}$
describes \cite{our} the quark mass dependence of the lattice
critical temperature at $\mu_B=0$  and the position of the phase
boundary   in the $(T,\mu_B)$--plane at  small $\mu_B$.
\begin{figure}[htb]
\begin{minipage}[t]{58mm} \hskip .3cm  \vskip -5.16cm
\includegraphics[width=14.5pc,height=12.pc,angle=0]{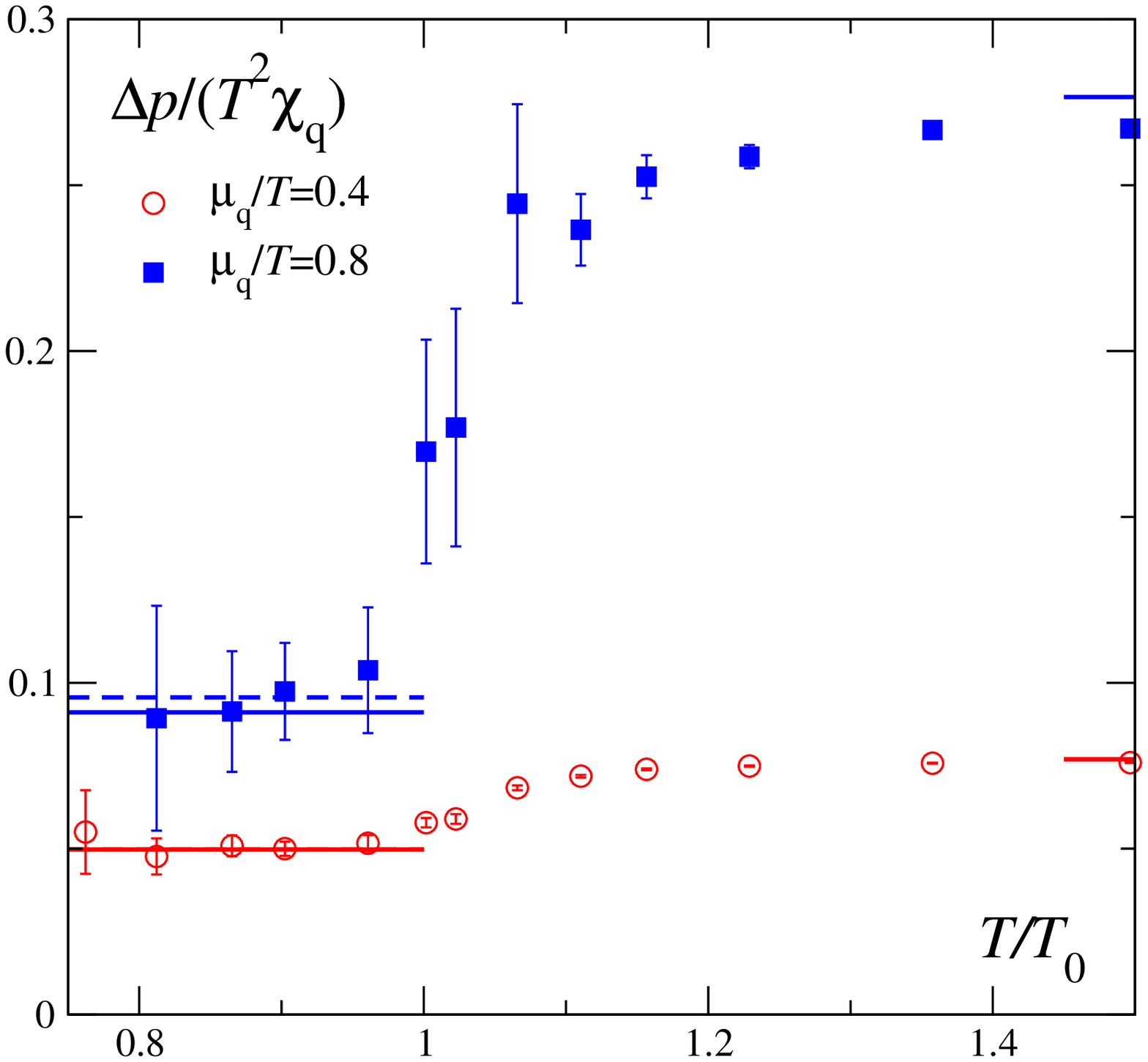}
\end{minipage}
 \hskip 0.5cm
 \begin{minipage}[t]{58mm}
{
\includegraphics[width=14.6pc, height=12.pc,angle=0]{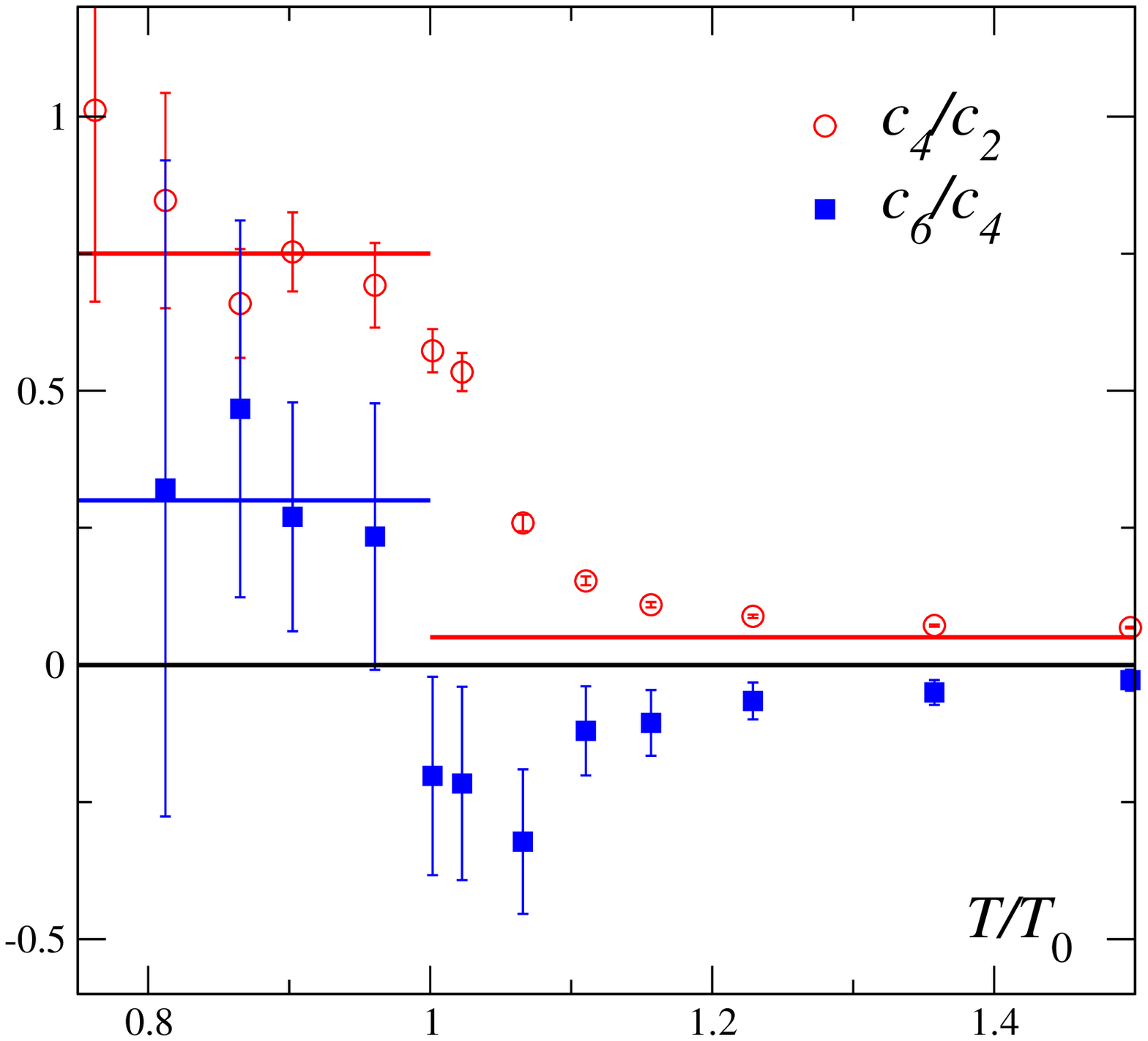}}\\
 \end{minipage} \vskip -0.5cm
\caption{ \label{fig:c2c4} Left--hand figure: temperature
dependence of $\Delta P/(\chi_qT^2)$ ratio  for two different
values of $\mu_q/T$. The 2--flavor  lattice results
 are from  \protect{\cite{lat}}. The lines are
the  hadron resonance gas model values  with (broken--line) and
without (full--line) Taylor expansion of $\cosh(\mu_B/T))$.   The
right--hand side shows the temperature dependence of the ratios of
second, forth and sixth order coefficients in the Taylor expansion
of thermodynamic pressure. The points are lattice results from
\protect{\cite{lat}}. The lines at $T<T_0$ are the values obtained
from the expansion of $\cosh(\mu_B/T))$  \protect{\cite{our}} .
  }
\end{figure}
\section{Resonances essential degrees of freedom near deconfinement}

The phenomenological partition function used in the description of
particles production in heavy ion collisions was, following
Hagedorn,  constructed as a non--interacting hadronic gas which is
composed of all hadrons and resonances. In the Boltzmann
approximation, suitable for the moderate values of $\mu_B<m_N$ and
$T\geq 50$ MeV, there is a factorization of $T$ and $\mu_B/T$
dependence in  relevant observables characterizing baryonic sector
of the system. The basic quantity is the pressure $\Delta P
=P(T,\mu_B)-P(T,\mu_B=0)$

\begin{equation}
\hspace*{-1.3cm}
{{\Delta P}\over {T^4}}=F(T)(\cosh ({{\mu_B}\over T}-1)~~~~~~{\rm
with}~~~~~~ F(T)\simeq \int dm\rho (m) ({m\over T})^2 K_2({m\over
T})
\end{equation}
from which $n_B$ and the baryon number
susceptibility $\Delta\chi_B$ are obtained as the
first and second order derivatives with respect to $\mu_B$, respectively.

The obvious consequences of the factorization in Eq.(1)  is that
any ratio of $n_B$, $\Delta P$ and $\Delta \chi_B$ at fixed
$\mu_B/T$ should be independent of $T$. This is the property which
can be directly checked with recent LGT results. In Fig.(1) we
show as an example the ratio of $\Delta P/\Delta (\chi_BT^2)$ for
two different values of $\mu_q/T$  as function of $T$.\footnote{
The quark chemical potential in Fig.(1), $\mu_q={1\over 3}\mu_B$}
It is clear from Fig.(1) that the factorization predicted by
$Z_{HG}$ is also seen in LGT results for  $T<T_0$, that is in the
confined phase of QCD. { \begin{figure}[htb]
\begin{minipage}[t]{58mm}
\hskip -1.3cm  \vskip -3.1cm \hskip -1.cm
\includegraphics[width=20.5pc,height=26.pc,angle=180]{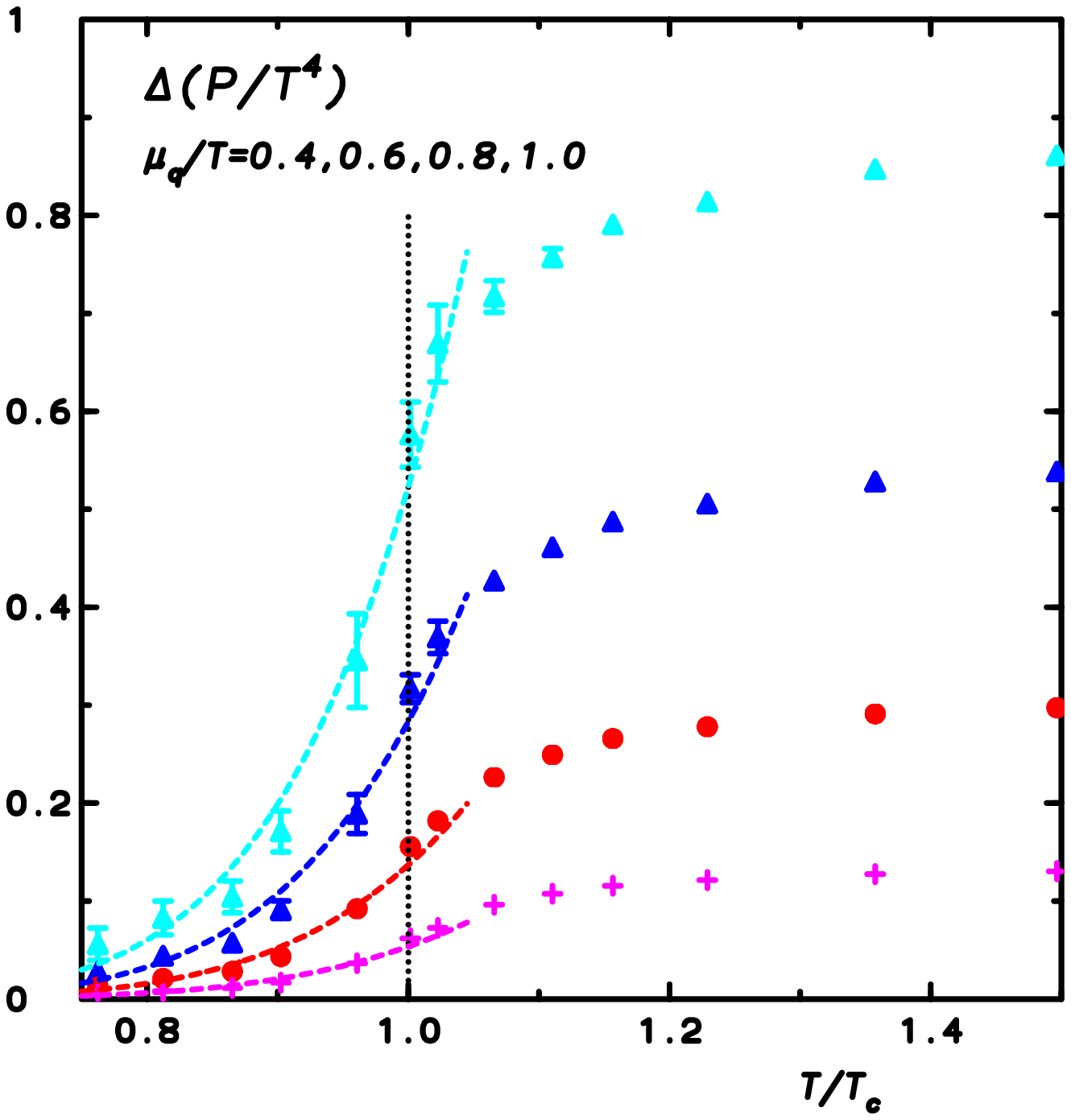}
 \end{minipage}
 \hskip 0.4cm
 \begin{minipage}[t]{58mm}
\vskip -1.1cm \hskip 0.3cm
\includegraphics[width=15.5pc,height=18.7pc,angle=0]{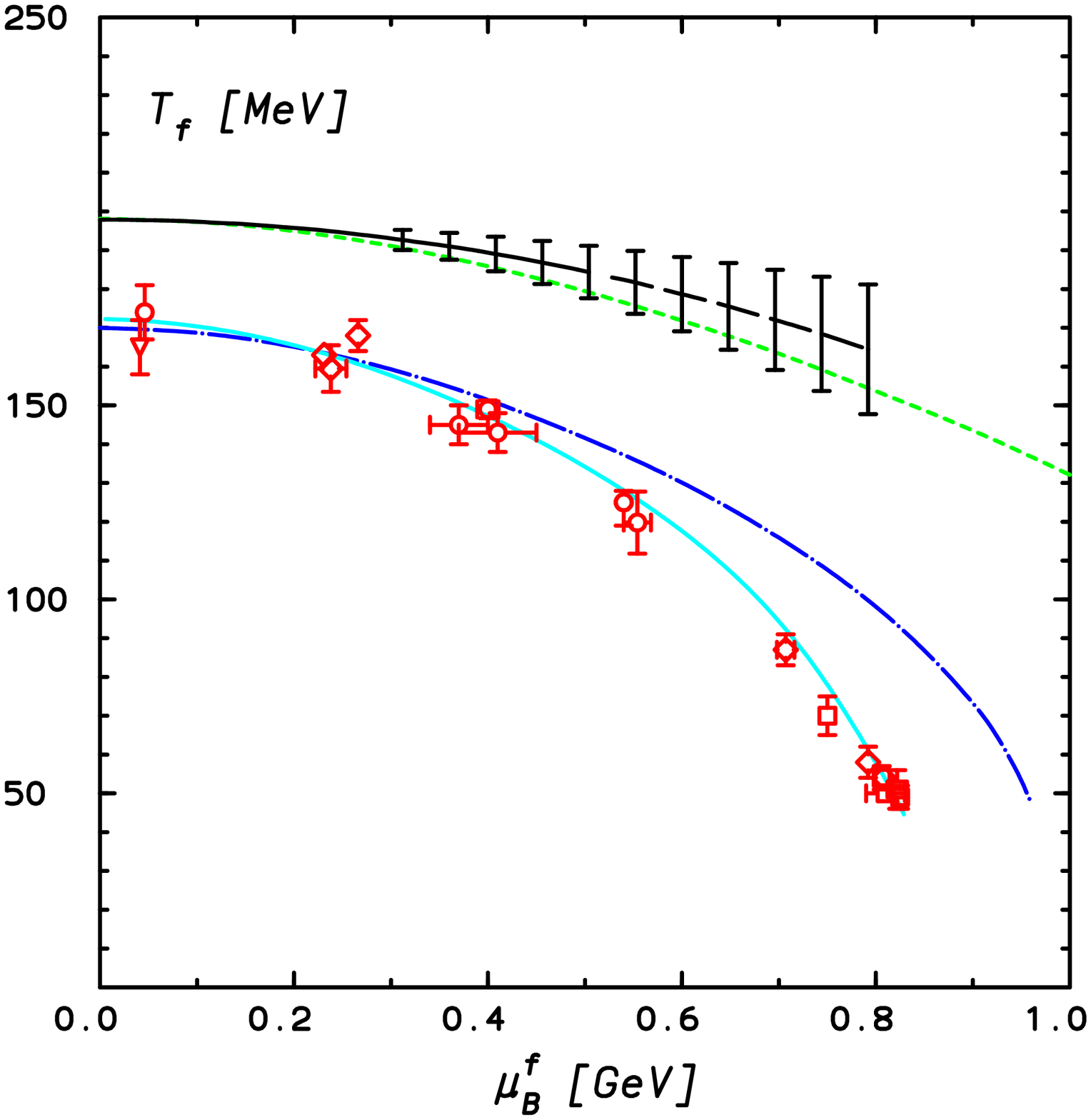}\\
 \end{minipage}
\vskip -2.6cm \caption{ \label{fig:c2c4}
 Left--hand figure: temperature dependence of baryonic pressure for different values of
$\mu_q/T$. The lattice results are from \protect{\cite{lat}}. The
lines are the statistical model results \protect{\cite{our}}. The
right--hand figure shows lattice results on phase boundary curve
(line with errors) together with phenomenological freeze-out
values of $T$ and $\mu_B$ (points) obtained from the analysis of
particle production  in heavy ion collisions \protect{\cite{rev}}.
Short-dashed and dashed--dotted lines are the statistical model
results obtained under the condition of fixed $\epsilon \simeq
0.6$ GeV/$fm^3$ with $m_\pi\simeq 0.77$ GeV and $m_\pi\simeq 0.14$
GeV respectively. Also shown (full--line) is the phenomenological
freeze-out curve of fixed energy/particle$\simeq 1$GeV from
\protect\cite{1gev}.
 }
 \end{figure} }

The second transparent feature of $Z_{HG}$ is that the $\mu_B/T$
dependence appears through a $\cosh$-function. The lattice results
at finite $\mu_q$ were obtained  using a Taylor expansion of the
pressure with respect to the quark chemical potential. So far, the
Monte Carlo results are available for the second $c_2$, fourth
$c_4$ and sixth $c_6$ order coefficient in the  Taylor series of
$P$ with respect to $\mu_q/T$. If the lattice thermodynamics was
consistent with the prediction of $Z_{HG}$ then one would expect
$c_2/c_4=3/4$ and $c_6/c_4=0.3$ being the coefficients in the
expansion of $\cosh(\mu_B/T)$ in Eq.(1). Fig.(1--right) shows the
corresponding ratios obtained on the lattice and the  results
obtained from Eq.(1). Within statistical error the agreement is
indeed seen to be  justified.

The pressure calculated on the lattice  (see Fig.(2--left) )
increases abruptly when approaching  deconfinement transition from
the hadronic side. If the phenomenological statistical operator
$Z_{HG}$ is of physical significance  then this increase could be
due to resonance formation. To check the importance of resonances
near deconfinement one would need to reproduce lattice results on
the $T$--dependence of $P$ at fixed $\mu_q/T$. However, to
quantify this dependence one needs to implement  the same set of
approximations in Eq.(1) as those being used on the lattice. First
of all current lattice results in Fig.(2--left) are obtained with
quite a large quark mass corresponding to $m_\pi\simeq 770$MeV.
This also distorts the baryon mass spectrum. Its pion mass
dependence can be deduced from lattice calculations at zero
temperature. We use the following ansatz for the parametrization
of the dependence of baryon masses on the pion mass
\cite{our,par},
\begin{equation}
{{m^*(m_\pi)}\over m}\simeq 1+A{{m_\pi }\over {m^2}}, \label{par}
\end{equation}
where $A=0.9\pm 0.1$,  $m^*$ is the distorted hadron mass at fixed
$m_\pi$ and $m$ is its corresponding physical value.

The lattice results were obtained in 2--flavor QCD, thus there is
no contribution of strange baryons in Eq.(1). Due to the Taylor
expansion of $P$ in  the lattice calculations one also needs to
perform a similar approximation in Eq.(1).  In Fig.(2-left) the
lattice results are compared with Eq.(1). The $T$ dependence of
QCD thermodynamics obtained on the lattice is seen in
Fig.(2--left) to be consistent with the predictions of $Z_{HG}$.

The value of $T_c$  was shown on the lattice to be dependent on
the pion mass \cite{pai}. The remarkable feature of this
dependence is that $\epsilon$  at $T_c$  is almost constant,
independent of $m_\pi$. This would suggest that deconfinement is
density driven and can be obtained from the condition of fixed
energy density. Fig.(2-right) shows recent lattice results on the
position of the phase boundary line in 2--flavor QCD within the
Taylor approximation and with the quark mass such  that
$m_\pi\simeq 770$MeV. The condition of fixed energy density
$\epsilon\simeq 0.6$ GeV/$fm^3$  with $\epsilon$ obtained from
Eq.(1) is seen to coincide with lattice results. Decreasing the
pion mass to its physical value and including complete set of
resonances expected in (2+1)--flavor QCD   results in the shift of
the position of the phase boundary line towards phenomenological
freeze-out condition of fixed energy/particle$\simeq 1$GeV.
 The splitting of freeze-out
and phase boundary line appears when the ratio of meson/baryon
multiplicities reaches the unity.

\section{Conclusions}
We have shown that the statistical operator of hadron resonance
gas used to  describe particle yields in heavy ion collisions
provide satisfactory description of recent lattice results on QCD
thermodynamics at finite chemical potential. In particular the
basic property  of this operator like e.g. factorization of
temperature and chemical potential dependence  is obviously
confirmed by the lattice results.   The ratios of the coefficients
in the Taylor expansion of thermodynamic pressure are also well
described by  the expansion of $\cosh$--function predicted by the
hadron resonance gas. These results are independent from the
particular choice  of the quark mass in the lattice calculations
and are also to large extent free from lattice artifacts.   The
phenomenological partition function was also  shown to describe
quantitatively the temperature and chemical potential dependence
of basic thermodynamical  observables bellow deconfinement. This
indicates that hadron resonance gas partition function is a  good
approximation of QCD thermodynamics of the hadronic phase. That is
why this partition function is also successful in heavy ion
phenomenology.

\section*{ Acknowledgments}
Stimulating  discussion with S. Ejiri is kindly acknowledged.
This work was partly supported by the  KBN under grant 2P03
(06925) and by the DFG under grant KA 1198/6-4.

   \end{document}